\documentclass[a4,preprint,authoryear]{elsarticle}
\biboptions{sort&compress}

\usepackage{acronym}
\usepackage{mdwlist}
\usepackage{amsmath}
\usepackage{amssymb}
\usepackage{prettyref}
\newrefformat{sec}{section~\ref{#1}}
\newrefformat{fig}{fig.~\ref{#1}}
\newrefformat{tab}{table~\ref{#1}}
\newrefformat{eq}{eq.~(\ref{#1})}
\newrefformat{apx}{\ref{#1}}

\usepackage{paralist}
\usepackage{epsfig}
\usepackage{verbatim}

\usepackage{natbib}
\bibpunct{(}{)}{,}{t}{,}{,} 

\usepackage{acronym}
\acrodef{ERC}{Extra chromosomal rDNA circle}
\acrodef{ODE}{Ordindary Differential Equation}
\acrodef{AF}{Ageing Factor}
\acrodef{AU}[AFU
]{Ageing Factor Unit}
\acrodef{DNA}{Deoxyribonucleic Acid}
\acrodef{rv}{random variable}

\DeclareMathOperator{\erf}{erf}
\DeclareMathOperator{\Var}{Var}
\DeclareMathOperator{\E}{\mathbb{E}}
\DeclareMathOperator{\const}{const}
\DeclareMathOperator{\Prob}{Pr}

\begin{document}

\title{The Accumulation Theory of Ageing} 

\author[UniS]{Andr\'e Gr\"uning\corref{cor}}
\ead{a.gruning@surrey.ac.uk}
\cortext[cor]{corresponding author. Tel.: +44-1483-68-2648; fax +44-1483-68-6051}
\author[UniS]{Aasis Vinayak PG}
\ead{p.vinayak@surrey.ac.uk}

\address{Department of Computing, University of Surrey, Guildford, GU2\,7XH, United Kingdom}


\begin{abstract}
Lifespan distributions of populations of quite diverse species such as humans
and yeast seem to surprisingly well follow the same empirical
Gompertz-Makeham law, which basically predicts an exponential increase
of mortality rate with age. This empirical law can for example be
grounded in reliability theory when individuals age through the random
failure of a number of redundant essential functional units. 
However, ageing and subsequent death can also be caused by the
accumulation of ``ageing factors'', for example noxious metabolic end
products or genetic anomalities, such as self-replicating
extra-chromosomal DNA in yeast.  

We first show how Gompertz-Makeham behaviour arises when ageing factor
accumulation follows a deterministic self-reinforcing process. We go
then on to demonstrate that such a deterministic process is a good
approximation of the underlying stochastic accumulation of ageing
factors where the stochastic model can also account for old-age
levelling off of mortality rate.


\end{abstract}

\maketitle




\section{Introduction}

Very simple organisms do not age, that is they do not show a
deterioration of significant life functions with age. Instead they
seem to simply break, when one of their essential functions happens to
fail. If such failures occur independently with a constant rate
$\lambda$ -- that is failures follow a Poisson process --  then their
corresponding survival function $S(t)$ decays exponentially with
time, and lifespans are distributed exponentially -- this is just a
simple conclusion from mathematical reliability theory and other
formally similar processes such as radioactive decay. 
\par
More complex organisms however do age, ie show a deterioration of life
functions, for example regarding regeneration after injury,
reproduction, metabolic activity and so on, and their mortality rates
$\lambda(t)$ tend to increase in time. For many different species, it
has been empirically established that mortality rates $\lambda(t)$
increase exponentially with time, following the famous empirical
Gompertz-Makeham law of mortality
\begin{equation}
  \lambda(t) = \lambda_0 + \alpha e^{\beta t}, \lambda_0, \alpha, 
  \beta > 0, \label{eq:gompertz}
\end{equation} 
at least approximately \citep{gavrilov:91} where $\alpha, \beta$ are constants and
$\lambda_0$ the age-independent death rate. Notable deviations are usually for very young
ages $t$ when ``childhood diseases'' lead to higher mortality and also
for old ages where empirically the mortality rate increases slower
than exponentially or even levels off to a constant value. 
\par
In this paper we will assume that the driver of ageing, and ultimately
death, is the accumulation of \acp{AF}, which for example plays a
major role in the ageing of yeast \citep{sinclair:98a}. We will demonstrate
generically that when such \acp{AF} are produced in an auto-catalytic 
or self-reinforcing process, then Gompertz-Makeham behaviour of the
mortality rate follows. Such self-reinforcing processes are quite
generic, for example in auto-catalytic chemical or biological
reactions or also in self-replicating genetic anomalities, and show
typically exponential behaviour. The term \emph{\acf{AF}} will
therefore generically stand for any substance or anomaly that is
thought to cause ageing in a specific organism. \acp{AF} could come in
discrete (even ``macroscopic'') chunks or be a continuous
quantity. They could be measured in absolute numbers of \acp{AU} or as
a concentration. 
\par
We proceed as follows: We will firstly take the naive point of view
that creation and replication of \acp{AF} is a deterministic 
process and mortality proportional to the abundance of \ac{AF}. This straight-forwardly leads to an exponentially increasing
mortality rate (with no old-age levelling). We then go an and assume that the creation and
replication of the \acp{AF} are stochastic when the \ac{AF} comes in
discrete units and in small numbers. If mortality rate is
assumed to be proportional to the probability that an organism has an
\ac{AF} abundance above a critical level, then it follows that the
mortality rate increases (more than) exponential for middle ages while
it levels off to a constant value at old-ages. 

\section{Background}

\subsection{Mortality Models} 

The above equation \prettyref{eq:gompertz} is primarily descriptive and empirically derived
from data, however does not in itself indicate a reason
\emph{why} mortality should increase exponentially with age (at
least for middle ages) or any underlying mechanism(s) of
ageing. However any such mechanism would need to be quite general since
lifespan of so many different species often surprisingly well fit
the Gompertz-Makeham law. Therefore various attempts have
been undertaken to ground Gompertz-Makeham behaviour in underlying
processes that could give an explanation as to why the mortality rate
should increase exponentially with time.  
\par
One prominent such attempt follows \citet{gavrilov:01}. These authors ground the
Gompertz-Makeham (and Weibull) laws in reliability theory. Their basic
assumption is that complex organisms have a block of redundant
functional elements for each 
essential function. Each such element does not age and hence fails with a constant rate. The
organism as a whole fails (or dies) if all redundant elements for a
single function have been exhausted. Depending on the number of
redundant elements and their failure state at birth, a Gompertz-Makeham
or Weibull law of mortality results. This model of ageing also nicely
explains the levelling off of the mortality rate at old ages. 
\par
A second attempt to ground the Gompertz-Makeham law in a biological
mechanism is \citet{shklovskii:05} starting from the assumption that
the ability of on organism to neutralise defective cells (or noxious
substances for that matter) is based on random encounters of these
harmful items with some neutralising antagonist. If the expected
number of such encounters sinks linearly with time, again the
Gompertz-Makeham law follows. However no attempt is made to explain
old-age longevity or to make plausible why the frequency of such random
encounters should decrease linearly in time.
\par
While the two models above are individual-based, ie when an individual
dies is determined by its life history, there is finally the
Penna model \citep{penna:95, laszkiewicz:05, stauffer:07}. This, too,
can explain an exponentially increasing mortality rate, albeit on a
population level. The Penna model assumes that genetic defects come to
bear at different ages, and individuals die if they have accumulated a
certain number of genetic defects -- in that sense it is similar to
ageing in yeast below. However how many defects have been acquired and
when they come to bear is predetermined genetically for each
individual, and only a population equilibrium of individuals with
different defective genes leads then to a Gompertzian exponentially
increasing mortality rate.

\subsection{Ageing in Yeast}

Our approach is motivated by systems that age through accumulating an
\ac{AF}. Yeast (\emph{Saccharomyces cerevesiae}) has in this respect been studied
extensively and served as model for ageing in higher
organisms \citep{sinclair:98a}. Albeit a single-cell organism, yeast
divides asymmetrically, so that mother and daughter cells can be
identified. Mothers eventually cease producing daughter cells, stall
and subsequently die. Extensive 
experimentation has shown that so-called \acp{ERC} are a probable
cause of ageing \citep{sinclair:97}. Finally, lifespan of yeast is
best measured in number of generations, that is how many times a mother
cell has already divided. While the absolute time between two such
divisions depends on the environmental conditions, lifespan distribution
measured in generational age is constant for different environments. 
\par
An \ac{ERC} is a snippet of a repetitive section of chromosomal
r\ac{DNA} that is excised from the chromosome spontaneously with a low
rate, and subsequently forms an extra-chromosomal \ac{DNA} 
ring. Excision can happen several times independently with approximately
constant low rate since the \ac{DNA} section in question contains about 100
repeats of the snippet. As the snippet has its own begin-of-replication
sequence, once excised it  replicates with a certain probability
in each cell cycle -- and hence replicates in a self-reinforcing process. 
\par
At cell division, the mother cell retains almost all of the
\acp{ERC} and does not share them -- eg proportional to cell volumes
-- with the daughter cells, so that \acp{ERC} accumulate in mother
cells whereas daughters usually start life free of \acp{ERC}. Cell
death then is assumed to occur because replication of a high number of
\acp{ERC} (in the order of $1000$) exhausts cell resources needed for
replication of the core chromosomal \ac{DNA}. 
\par
For very old mother cells \ac{ERC} retention capability seems to 
saturate, so that \acp{ERC} are shared with daughters, effectively
decreasing the mother's \ac{ERC} replication rate and producing
prematurely aged daughters with a reduced life span. Modelling the
effects of retention saturation is beyond the scope of this
paper. From figures discussed in the literature \citep{sinclair:97,
  sinclair:98, gillespie:04}, it seems however that the reduction in
\ac{ERC} number is minor compared to their absolute number and/or rate
of replication, so that mothers' lifespans are extended around one or
two generations at most. It does however significantly reduce the
expected lifespan of daughters of old mothers that have received a
portion of the mother's \acp{ERC}. The model discussed below can
straight-forwardly  be modified to take into account birth with a
number of \ac{ERC} but for sake of simplicity we will only deal with
populations that start life from a clean \ac{ERC}-free state as has
been the case in many experiments \citep{sinclair:97}.
\par
Experiments  suggest as well that for yeast, cells live on
quite happily when they have low or medium numbers of \acp{ERC} but die when they have reached a 
critical level of \acp{ERC} so that there does not seem to be a linear 
increase of mortality rate with number of \acp{ERC} as we will naively
assume in the deterministic model. Instead the underlying
assumption seems to be that mortality shows an increase at or around a
critical number of \acp{ERC} \citep{sinclair:98}. 
\par
To conclude, ageing (in yeast and other organisms,
\citealt*{sinclair:98a}) can be caused by an \ac{AF} that comes
in discrete chunks and where individuals initially start up with no or a
very small number of such \ac{AF}. The number of such \ac{AF}
increases due to two processes: initial creation of such \acp{AF}
(in yeast by excision from chromosomal \ac{DNA}) and subsequently
self-reinforcing replication (for yeast once in the cell
cycle). Finally mortality is correlated with the abundance of such
\acp{AF}. 
\par
In the next section we will present two formalisations of these
processes. The first is naive, straight-forward and deterministic. The  
second however is more complicated and based on stochastic processes.
This stochastic model serves also to justify the  naive
approach. These approaches are certainly motivated by the experimental 
data for yeast, but we will formulate them quite generally as
auto-catalytic processes are an ubiquitous phenomenon in nature, thus
might well play a role in ageing processes in many organisms.

\section{Models}

Let $c(t)$ denote the abundance of an \ac{AF} at time $t$. The
abundance could either stand for the (discrete) absolute number of
\acfp{AU} if the \ac{AF} comes in discrete chunks and absolute numbers are low,
or for a (continuous) concentration if absolute numbers are high or
the factor itself is a continuous variable. We shall assume that $c(t)$
changes in time due to two processes:
\begin{inparaenum}
\item Creation with rate $p(t)$, ie new \acp{AU} come into existence
  independently of existing ones. This could be from internal processes
  such as a product of a chemical or metabolic reaction or the
  excision of snippets of \ac{DNA} as in yeast, but
  also external such as a \ac{DNA} defects due to UV light and so
  on. 
\item Replication with rate $r(t)$, ie once some \acp{AU} have been
  acquired there is an auto-catalytic process that produces more
  \acp{AU}, linearly depending on the abundance $c(t)$ of \ac{AF}
  already present, ie $\Delta c(t) \sim c(t)$. Typical examples of such processes are
  auto-catalytic chemical reactions where the product at the same time
  is also an educt of the reaction, or the replication of a \ac{DNA}
  snippet independently of the core genome as happens with the \acp{ERC} in
  yeast. 
\end{inparaenum}
These processes could be happening in continuous or discrete time, on
an absolute time scale or tied to an internal time scale. For yeast
for example, \ac{ERC} replication is synchronised with the cell cycle.
\par

\subsection{Deterministic Production of \acp{AF}} \label{sec:deterministic}

As it is simpler and instructive, we first consider the case where the \ac{AF}
comes in large quantities in each individual and creation and
replication processes have high rates. This means that while
individual creation and replication may well be stochastic, we can
assume -- due to the high rates and high numbers of \ac{AF} involved --
that creation and replication proceed deterministically, and
variations of \ac{AF} around its instantaneous mean are negligible. 
\par
We will in addition assume continuous time. Mathematical argument is
often easier in continuous time than in discrete time, and the
inaccuracies incurred by transforming equations between discrete time
and continuous time are small when the involved quantities are
smooth. 
\par
Formalising creation and replication processes in continuous time, $c(t)$ follows an
inhomogeneous linear \ac{ODE}:   
\begin{equation}
  \frac{dc(t)}{dt} = r(t) c(t) + p(t) \label{eq:ode0}
\end{equation} with solution for initial value $c(0) = c_0$ (see \prettyref{apx:linode}):
\begin{equation}
  c(t) = \int_0^t p(\tau) e^{\int_\tau^t r(\tau') d \tau'} d\tau  +
  c_0 e^{\int_0^tr(\tau)d\tau} \label{eq:sol0} 
\end{equation}
$c(t), p(t), r(t)$ are non-negative functions, and it is obvious that
$c(t)$ increases essentially exponentially with $t$ when $r(t) >
0$. For discrete time, $c(t)$ would be the solution of a similar
difference equation which again increases exponentially
(geometrically) in time. 
\par
For simplicity let us assume $r(t) = r$ and $p(t) = p$ are constant,
so that both creation and replication processes of \ac{AF} do
not change with time, and \prettyref{eq:sol0} simplifies to
\begin{equation}
  c(t) = \frac{p}{r} (e^{rt} - 1) + c_0 e^{rt} \label{eq:c}  
\end{equation}
However time-varying $r(t)$ and $p(t)$ could also easily be
accommodated and would not essentially change the arguments belows. 
\par
If we assume the death is only due to the \ac{AF}, then a natural
assumption is that the mortality rate is proportional to $c(t)$, ie:
\begin{equation}
  \lambda(t) = z c(t), z > 0 \label{eq:lambda}
\end{equation} where $z$ is the proportionality factor. 
Then quite straight-forwardly, inserting \prettyref{eq:c} into
\prettyref{eq:lambda} provides a Gompertzian time development of $t$ with
some additional terms to adapt to initial conditions. For higher $t$,
$\lambda(t)$ increases exponentially with $t$, and does not level off
to a constant value either for very high $t$.

\subsection{Example: Lifespan of Yeast}

\begin{figure}
  \centering
  \epsfig{width=0.6\hsize,file=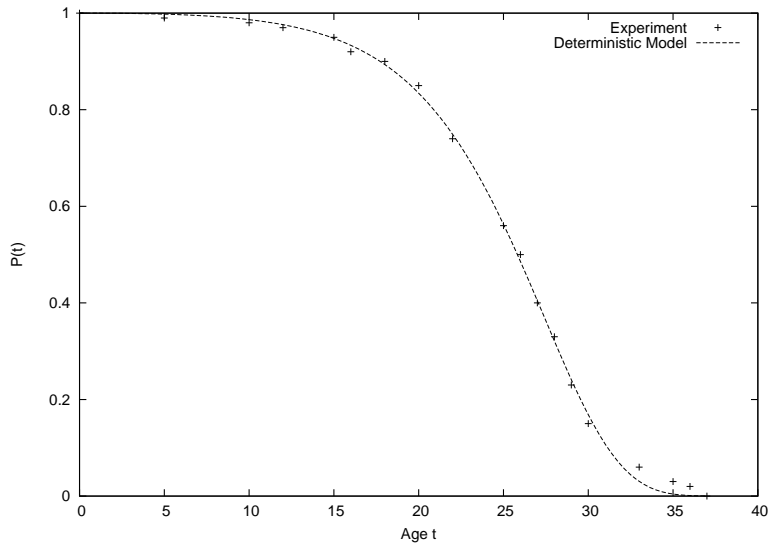,angle=-90}
  \caption{Fitting the survival curve of yeast: A least squares fit
    for $p':=zp$ and $r$ of \prettyref{eq:det} against
    \citeauthor{sinclair:98}'s data
    \citeyear{sinclair:98} (Fig.~1). $c_0$ was set to $0$, as
    experiments started with \ac{ERC} clean cells. Also no
    age-independent mortality rate $\lambda_0$ was used, as accidental
    death is often negligible under
    lab-conditions. Best least squares fit for $p' = 0.00011, r =
    0.22$, $SSR =  0.0034$,  
    $RMS =  0.014$. Note the (expected) underestimation of old-age
    longevity.} \label{fig:sinclair}
\end{figure}

Experimental data is often in the form of lifespan distributions
rather than mortality rate per se. The lifespan distribution is given
by the survival function $S(t)$ which is the probability that an individual
survives at least until $t$. It is related to the mortality rate
$\lambda(t)$ via the following \ac{ODE} with initial condition $S(0) =
1$: 
\begin{equation}
  \frac{dS(t)}{dt} = -\lambda(t) S(t).
\end{equation}
This is again of the form as in \prettyref{apx:linode} and
inserting \prettyref{eq:c} and \prettyref{eq:lambda} and solving for
$S(t)$ (see \prettyref{eq:P}) we get:
\begin{gather}
  S(t) 
  = e^{-\int_0^t \lambda(\tau) d\tau} 
  = e^{-\frac{zp}{r} \left[\frac{e^{rt}  - 1}{r} -t \right]} 
  e^{-\frac{zc_0}{r}  \left[ e^{rt} -1 \right]}. \label{eq:det} 
\end{gather}
This form of $S(t)$ can now straight-forwardly be fitted to
experimental data, see \prettyref{fig:sinclair}.
\par
\par
As $p$ and $c_0$ do not enter \prettyref{eq:det} independently of $z$,
their absolute value cannot be estimated from data unless further biological assumptions are taken into account. $r$
can be seen as the  average or effective rate of increase of
the \ac{AF}. The fit is surprisingly good  even though it does not
take into account the levelling of the mortality rate $\lambda(t)$
observed for many species for high $t$ so it underestimates the frequency of very high lifespans.

\subsection{Stochastic Production of \acp{AF}}

While \prettyref{eq:det} often yields a good fit as we have seen, it
starts from the assumption that the \ac{AF} comes in high numbers with
high rates $p$ and $r$, so that the process of \ac{AF} is largely
deterministic. The deterministic model does also not account for the
levelling off of mortality rate with time.   
For example for yeast it is however known that the \ac{AF} comes only
in discrete units and individuals acquire them one by one, and 
numbers stay smallish throughout life-time. All this
does not warrant a deterministic approach. 
Hence for such discrete \ac{AF}s in small numbers,
creation and replication should be treated as \emph{stochastic
  processes}. For simplicity of argument here, we will assume discrete
time in the following.
\par
The stochastic creation process of discrete
\acp{AU} is easily modelled as a (constant or varying rate) Poisson or
Bernoulli process, so for constant rate $p$ the expected number of \ac{AU}
increases as $pt$ with variance $pt$. In the following we will assume
that $p$ is constant. This is simpler and there is also no biological
indication to contrary. 
\par
The replication process is
more complicated and best be described in the framework of branching
processes \citep{haccou:05}, a type of stochastic process suited to
model growth processes. We will first look at a pure branching process
that corresponds to the stochastic \ac{AU} replication in isolation,
and subsequently combine replication and creation processes. 
\par
In the following capital letters such as $X$ denote random
variables. $\Prob(\cdot)$ denotes the probability (density) of an event,
$\E(\cdot)$ and $\Var(\cdot)$ the expectation value and variance of a
random variable respectively. For further background -- such as
relations of conditional expectations and variances or approximations of
the Q- and $\erf$-function -- we refer the reader to good textbooks with an introduction to
probability theory, for example \citet{cover:91} or \citet{haccou:05}. 

\subsection{Replication Process Only} \label{sec:replication}

In this section we concentrate on the replication process per se, and 
assume that initially one \ac{AU} is already present: $c(0) = 1$. Then
existing \acp{AU} replicate with mean rate $r$ in each time step. This
means each single \ac{AU} at $t$ gives rise to one (itself) or two
\acp{AU} (itself and its replication) at $t+1$ with probabilities $1-r$ and $r$
respectively. Formally individual replication is a random variable
$\Xi$ with probabilities $\Prob(\Xi = 1) 
= 1-r$ and $\Prob(\Xi = 2) = r$, and all other outcomes have
probability zero, and we define short symbols for mean and variance of
the number of successors of a single \ac{AU}: $m := \E(\Xi) = 1+r$ and
$s^2 := Var(\Xi) = r(1-r)$. 
\par
As the replication process itself is random, also the abundance of the
population $c(t)$ at time $t$ is a random variable and -- to mark that
the abundance is now a random variable in discrete time --  we replace
symbol $c(t)$ with $C_t$ and call it the population size.    
\par
In fact, the sequence of $(C_t)$ forms a \emph{branching process}
(see \prettyref{apx:branch}) and $C_{t}$ is the $C_{t-1}$-fold sum of the
individual replication process $\Xi$ (with i.i.d.\ outcomes; we refrain
from distinguishing the individual $\Xi$ in order not to overburden
the notation): $C_t = \sum_{i = 1}^{C_{t-1}} \Xi$. The expectation
value $\mu(t):= \E(\Xi)$ and variance $\sigma^2(t):=Var(C_t)$ of the population sizes
$C_t$ then follow from expectation value and variance of individual
offspring $\Xi$ as (see \prettyref{apx:branch}):
\begin{equation}  
  \mu(t) = m^t, \quad
  \sigma^2(t) = s^2 \frac{m^t (m^t - 1)}{m(m-1)}
\end{equation}
with the initial condition that we start from a single \ac{AU} at $t =
0$, ie $C_0 = 1$.
\par
In the case of deterministic growth, the mortality rate $\lambda(t)$
was assumed proportional to $c(t)$. We could in principle also here
set $\lambda(t) \sim \E(C_t) = m^t$ with the same results as above,
loosing again the levelling off of $\lambda(t)$ for old ages. 
Experiments indicate that for yeast, cells with a relative
low number of \acp{ERC} do not have an altered instantaneous mortality
rate from cells with no \acp{ERC}. The mortality rate  
seems to increase only when the number of \acp{ERC} has reached a
critical level. Hence in this stochastic process setting, it is more natural to
assume that the mortality rate $\lambda(t)$ is proportional to the
probability that an individual has more \ac{AU}s $C_t$ than a fixed
critical level $c_\dagger$, ie  
\begin{equation}
  \lambda(t) = z \Prob(C_t \geq c_\dagger), z > 0.
\end{equation} 
This agrees also with the type of mortality criteria used in computer
simulations of ageing processes \citep{gillespie:04}. More 
complicated dependencies are of course possible, but would typically
show a similar dependency on an upper quantile of $C_t$.  
\par
In order to calculate $\lambda(t)$ explicitly, we would need to know
the distribution of $C_t$ for all times $t$. To our knowledge this
problem has not been solved explicitly. We can however make a series of
approximations for small times $t$, large times $t$ and middle times
$t$ to demonstrate that this setting for the mortality rate
$\lambda(t)$ yields Gompertz-Makeham behaviour for middle $t$ and a
constant mortality for very old ages.
\par

\paragraph{Small $t$}

We assume that at time $t = 0$ there is one \ac{AU}, $C_0 = 1$. As
each \ac{AU} can have at most two successors (itself and its
offspring), we know that for $ t   < log(c_\dagger) /  log(2)$,
$\Prob(C_t \geq c_0) = 0$. So depending on $c_\dagger$ there will be more or less
long an initial period for which the mortality rate is zero. 

\paragraph{Approximation with normal distribution} 

Generally the precise form of $\Prob(C_t)$  is not
known. Knowing only its $\mu(t)$ and $\sigma(t)$ we
assume that  $C_t$ follows approximately a normal distribution with mean
$\mu(t)$ and variance $\sigma^2(t)$. This is plausible as $C_t$
ultimately in our setting is a sum of binomial distributions (each
individual either replicates or not and survives itself into the next
generation, so $C_{t+1} = C_t + B(C_t, r)$)  which  themselves can be
approximated by a normal distribution -- however the $C_t$ are not independent. Also quantities related to $C_t$ such as $\Prob(C_t|C_{t-1})$ 
have been shown to approximate a normal distribution
\citep{quine:94}. Hence for mildly large $t$ we approximate: 
\begin{equation}
  \Prob(C_t \geq c_\dagger) \approx  Q\left(\frac{c_\dagger - \mu(t)}{\sigma(t)}\right) =
  \frac{1}{2} - \frac{1}{2} \erf\left(\frac{c_\dagger - \mu(t)}{\sqrt{2} \sigma(t)}\right) \label{eq:approx} 
\end{equation} where $Q(x) = {1}/{2} - {1}/{2} \erf(x/\sqrt(2))$ is the
Q-function that gives the mass of a standard normal
distribution above $x$ (ie the probability of getting a value larger
than $x$).     

\paragraph{Large $t \gg 0$}

For large very large $t \gg 0$, 
\begin{equation} 
  \lim_{t \to \infty} \frac{c_\dagger - \mu(t)}{\sigma(t)} 
  = \lim_{t \to \infty} \frac{c_\dagger - m^t}{s \sqrt{\frac{m^t ( m^t
        -1)}{m(m-1)}}} = -\frac{1}{s}   
\end{equation}
as both $\mu(t), \sigma(t)$ grow exponentially with the same
rate. Hence 
\begin{equation}
  \lim_{t \to \infty} \lambda(t) = Q(-1/s) > 0,
\end{equation}
ie the mortality rate $\lambda(t)$ levels off with age to a constant
positive value, hence $\lambda(t)$ shows the experimentally observed
old-age deviation from the Gompertz-Makeham law.

\paragraph{Middle $t$}

We have established that $\lambda(t) = 0$ for small $t$ and $ \approx \const
> 0$ for large $t$. As each \ac{AU} survives into the next time step,
$C_{t+1} \geq C_{t}$, ie the number of \acp{AU} is
non-decreasing. This means that $\lambda(t) = z \Prob(C_t \geq c_\dagger)$
is non-decreasing as well. Hence $\lambda(t)$ has roughly the form of a
(discrete) sigmoid.
\par
If approximation of $C_t$ with a normal distribution is valid also for 
middle $t$ such that $0 \ll \mu(t) < c_\dagger$, where $P(C_t >
c_\dagger) > 0$ but the expected number $\mu(t) = m^t$ still less than the
critical value $c_\dagger$, we go on to show 
that increase of $\lambda(t)$ must be steeper than $m^t$, in other words we
have an exponential growth of the mortality rate for intermediate
ages.
\par
If $x:= {c_\dagger - \mu(t)} / {\sigma}$ is not too close to zero, ie
$\mu(t)$ still much smaller than $c_\dagger$, then $Q(x) \approx
\frac{1}{\sqrt{2\pi} x} e^{-\frac{1}{2}  x^2}$  for $x > 0$ and
hence 
\begin{equation} 
  \lambda(t) 
  \approx Q\left(\frac{c_\dagger - \mu(y)}{\sigma(t)} \right) 
  \approx \frac{\sigma^t}{\sqrt{2\pi} (c_\dagger - \mu(t))} e^{-\frac{(c_\dagger -
      \mu(t))^2}{\sigma^2(t)}}. 
\end{equation}
For $c_\dagger > \mu(t)$, the exponential is increasing towards $1$. The
denominator of the pre-factor is decreasing, and its numerator
increasing exponentially approximately as $m^t$ (since 
$\sqrt{\frac{m^t(m^t -1)}{m(m-1)}} \approx m^t$ for not too small
  $t$). Hence there are intermediate $t$  for which $\lambda(t)$
  approximately increases more than exponentially:  
  \begin{equation} 
    \lambda(t) \gtrsim z m^t 
\end{equation}

\subsection{Creation and Replication} \label{sec:full}

Unless born out of very old mother cells, new-born yeast cell usually
do not have any \ac{ERC}, ie $C_t = 0$ so creation and replication
process need to interact to cause ageing. Before replication of \acp{AU}
can start, first some must be produced, ie we have a combination of
two stochastic processes. This combination is more cumbersome to deal
with, but we 
demonstrate below that (asymptotic) growth rates of $\E(C_t)$ and
$\Var(C_t)$ are unaltered, so that the same approximations and
asymptotic considerations are valid for this more complicated process
as were in the case of replication only.
\par
Let $C_t$ denote the \ac{AU} population size as before, and $\Xi$
again the individual \ac{AU} replication process with the same
distribution. Finally let $X_t$ be the creation process with
constant rate $p$, ie $\Prob(X_t = 0) = 1-p$ and $\Prob(X_t = 1) = p$. Then
$\E(X) = p$ and $\Var(X) = p(1-p)$.
\par
The random number $C_{t+1}$ of \ac{AU} at time $t+1$ is then the sum of the
replication of the previous number of \acp{AU} $\sum_{i=1}^{C_{t}}
\Xi$ plus the number $X_t$ of newly created \acp{AU}:
\begin{equation}
  C_{t+1} = \sum_{i=1}^{C_{t}} \Xi + X_t \label{eq:sum}
\end{equation}
Utilising independence of the individual $\Xi$ and $X_t$ we can
calculate a recursive equation for $\E(C_t)$ as in the case above, see
\prettyref{apx:full}:
\begin{equation}
  \E(C_{t+1}) = m \E(C_t) + \E(X_t) \label{eq:recurrent} \\ 
\end{equation}
with explicit solution for initial value $C_0 = 0$:
\begin{equation}
  \E(C_{t}) = \frac{p}{m-1} (m^t -1). \label{eq:meansol} \\ 
\end{equation} 
Comparison of \eqref{eq:recurrent} and \eqref{eq:meansol}
with \eqref{eq:ode0} and \eqref{eq:c} also makes it clear that the
process $C_t$ is the stochastic time-discrete analogue to the
deterministic continuous time equation in
\prettyref{sec:deterministic} (with $r = \log(m)$). 
\par
Likewise a recursive formula for $\Var(C_t)$ can be  obtained: 
\begin{equation}
  \Var(C_{t+1}) = m^2 \Var(C_{t}) + s^2 \E(C_{t}) +
  \Var(X_{t}). \label{eq:var} 
\end{equation}
This has a rather tedious explicit solution, see
\prettyref{apx:full}. For large $t$ however it is clear from the
dependency of $\Var(C_t)$ on $\Var(C_{t-1})$ and because $\E(C_t)$
only grows like $m^t$ for large $t$, that the growth of $\Var(C_t)$
will 
asymptotically be like $m^{2t}$. Ie both $\Var(C_t)$ and $\E(C_t)$ grow
asymptotically as in the case before. Hence again under the assumption
that $C_t$ can be approximated well enough with a normal distribution,
the same considerations as above for small, middle and 
large follow, ie $\lambda(t) = 0$ for small $t$, $\lambda \gtrsim
zm^t$ for middle $t$ and $\lambda(t)$ constant for large $t$. 

\section{Discussion}

The production processes and death criteria considered in this paper
arise from experimental observation of ageing in yeast mother cells
where \acp{ERC} seem to be drivers of ageing and death. It is
remarkable indeed that the deterministic equation \prettyref{eq:det}
already yields such a good fit with experimental data, even though the
underlying 
process model is deterministic, uses a non-biological death criterion
and finally does not take any detailed properties of the underlying
\ac{AF} production processes into account such as sharing of \acp{ERC} between
old mothers and daughters. It only relies on the exponential
accumulation of \acp{AF}. The model does not take into account either
that ageing in yeast need not necessarily only take place through 
accumulation of \ac{ERC}. 
\par
In the stochastic model, we have demonstrated that the mortality rate
increases with time roughly like a discrete sigmoid and $\lambda(t)
\gtrsim zm^t = ze^{rt}, r = log(m)$ for middle $t$. The good fit
of an exponential mortality rate with data (eg in
\prettyref{fig:sinclair}) shows that usually 
$\lambda(t)$ does not increase significantly faster than exponentially
or that such faster increase has no significant effect on the form of
the survival function. If we do not care about the overestimation of
mortality old ages, we can extend the exponential approximation of
$\lambda(t)$ also to greater $t$. Finally we know that $\Prob(C_t \geq
c_\dagger) = 0$ strictly for young ages. Extending the approximation
of $C_t$ with a normal distribution also to these young ages would
lead to a $\Prob(C_t \geq c_\dagger) > 0$. However as then also
$c_\dagger \gg m(t)$ 
this is often negligibly small (and would for example be masked by a
constant age-independent death rate $\lambda_0$). Hence the
deterministic approach $\lambda(t) \sim e^{rt}$ can be justified for all
ages $t$ as an approximation of the stochastic approach $\lambda(t) \sim
\Prob(C_t \geq c_\dagger)$.
\par
\citet{gillespie:04} set out to understand ageing in yeast with the help of numeric computer
simulations. Essentially they simulated a stochastic process such as
in \prettyref{sec:full}. Their simulation also modelled the old-age
\ac{ERC} sharing between mothers and daughters. Simulation parameters,
such as replication rate $r$ or excision rate $p$ were estimated from
data in the literature. With these estimated parameters, their
simulation yielded a fairly good fit which however was not
quantified. The best fit was achieved by assuming that $p(t) \sim
t^2$, ie a non-constant creation rate which seems to be an ad-hoc choice
to improve the fit over a constant excision rate. 
In light of the good fit of
\prettyref{eq:det} which was derived with a constant $p$, it is quite
surprising that \citeauthor{gillespie:04} did not obtain a good fit
under the same assumption.   
\par
Finally the rate with which \acp{ERC} in yeast are  
replicated has been estimated as about $m = 1.6$ \citep{sinclair:97}
-- this follows as an average rate from an average lifespan of 15
generations once the first \ac{ERC} has been acquired and a critical
level in the order of $1000$, as $m \approx 1000^{1/15} = 1.585$
and hence $r = \log 1.59 \approx 0.46$, this is about twice the $r$
estimated from the data in \prettyref{fig:sinclair}. 
\par
Last but not least, \prettyref{eq:det} and its good fit to the data also
confirm  quantitatively a qualitative conjecture of \citet{sinclair:98}, namely
that the survival function for yeast is a ``sum'' of a decay process (due
to excision) and subsequent exponential accumulation of \acp{ERC},
albeit the combination of these two process is a more involved sum
than \citeauthor{sinclair:98} implied, namely that of two stochastic
processes in \prettyref{eq:sum}.

\section{Conclusion}

In this paper we have demonstrated that mortality rates similar to
those in the Gompertz-Makeham law can be derived when we assume an
\ac{AF} is produced in an auto-catalytic process. Such processes
in turn lead to an exponential (or synonymously geometric) increase of
the abundance of \acp{AF}. In the  case of deterministic production of
\acp{AF}, with 
mortality assumed proportional to the \ac{AF} abundance a plain
Gompertz-Makeham law follows with the usual underestimation of
longevity for old ages. In the stochastic case, not only does a
Gompertz-Makeham law follow, but also the levelling off of mortality
rates is derived if mortality is assumed proportional to the
probability that an individual has reached a critical level of
\acp{AF}.  
\par
All in all the paper establishes a useful and interesting relation
between mortality rates that follow a Gompertz-Makeham law and an
underlying auto-catalytic process. For yeast ageing driven by \acp{ERC}
the model and data are in qualitative agreement. However discrepancies
between the rate of replication estimated here and within other
experimental contexts need to be examined. 
\par
The derived relation is general and applicable to all organisms where
ageing is based on exponential accumulation of an \ac{AF} -- this
nicely contrasts and complements 
\citeauthor{gavrilov:01}'s approach (\citeyear{gavrilov:01}) for which
ageing is based on the exhaustion of parallel redundant functional elements.

\section*{Acknowledgments}

We thank Colin S.\ Gillespie for making details of simulations in
\citet{gillespie:04} available to us. 
Vinayak PG is funded by a PhD studentship of the Department of Computing
at the University of Surrey.

\begin{appendix}
  
  \section{General Linear \ac{ODE}} \label{apx:linode}

In the present paper the linear inhomogeneous \ac{ODE} plays are role: 
\begin{equation}
  \frac{dx(t)}{dt} = a(t)x(t) + b(t) \label{eq:linode}
\end{equation} where $a(t)$ can be thought of as a time-variant rate of
growth of $x(t)$ and $b(t)$ as a time-variant external influx to
$x(t)$. The generic solution can be obtained -- under some smoothness
assumptions -- for example using the Green's function approach. We
refer the reader to a generic textbook on \acp{ODE} and just state the
generic solution below. The reader can check the correctness of the
solution through differentiating with respect to $t$ and inserting into
\prettyref{eq:linode}. 
\par
In the homogeneous case, i.e. $b(t) = 0$, the general solution of \prettyref{eq:linode} with initial
value $x(0) = x_0$ is 
\begin{equation}
  x(t) = x_0 e^{\int_0^t a(\tau) d\tau}.
\end{equation}
In the inhomogeneous case $b(t) \neq 0$, the general solution with
initial value $x(t) = x_0$ is:
\begin{equation}
  x(t) = \int_0^t p(\tau) e^{\int_\tau^t r(\tau') d \tau'} d\tau  + x_0 e^{\int_0^tr(\tau)d\tau} 
\end{equation}
For example if $a(t) = a$, $b(t) = b$ are constant and $x(0) = 0$ the
solution is: 
\begin{equation}
  x(t) = \frac{b}{a} (e^{at} -1)
\end{equation}

\section{Lifespans and mortality rate} 

Let $\lambda(t)$ be the instantaneous mortality (or failure) rate, ie
the rate with which an individual dies/fails at $t$, then is it
related to the distribution of lifespans or survival function $S(t)$ (ie the
probability for an individual to survive until time $t$) via:
\begin{equation}
  \frac{dS(t)}{dt} = -\lambda(t) S(t)
\end{equation}
with initial condition $S(0) = 1$. This equation is of the form of
\prettyref{eq:linode}, hence solved for $S(t)$ we obtain:
\begin{equation}
  S(t) = e^{-\int_0^t \lambda(\tau) d\tau}. \label{eq:P}
\end{equation}
\par
If we have independent processes that cause death with rates
$\lambda_1(t)$ and $\lambda_2(t)$ and corresponding survival functions
$S_1(t), S_2(t)$ and combine them as $\lambda(t) = c_1
\lambda_1(t) + c_2 \lambda_2(t)$, then due to linearity of integration
and because the exponential transfers sums into products: 
\begin{equation}
  S(t) = S_1^{c_1}(t) S_2^{c_2}(t).
\end{equation}
This is handy when dealing with different terms in a complex
$\lambda(t)$ separately. For example adding constant age independent
mortality rate $\lambda_0$ to a $\lambda(t)$ means that the
corresponding $S(t)$ acquires an additional factor $e^{-\lambda_0t}$.  

\section{Branching Processes} \label{apx:branch}

Intuitively a branching processes is a model for stochastic
reproduction of individuals in discrete time and follows the
population size $C_t$ in time. Each individual lives for one time step
(or generation), and produces a number of offspring according to a
\ac{rv} $\Xi$ with mean of $m:=\E(\Xi)$ and variance $s^2:=\Var(\Xi)$
according to a fixed probability distribution.   
Given the population size $C_t$, the number of individuals in the next
generation $C_{t+1}$ is a random number that is the sum of all
offsprings produced randomly and independently by all individuals in
$C_t$, ie $C_{t+1} = \sum_{i=0}^{C_t} \Xi$. The sequence of $(C_t)_{t
  \geq 0}$ forms a stochastic process, ie a sequence of interdependent
random variables \citep{haccou:05}.   
\par
Let $\E(X)$ denote the expectation value of a random variable $X$ and
$\Prob(X)$ and $\Var(X)$ its probability and
variance. For given expectation $m$ of an individual's 
offspring and initial population $C_0 = 1$, one can then calculate the
expectation values $\mu(t):= \E(C_t)$ of expected population
sizes. Since $\E(C_{t+1}|C_{t}) = \sum_{i=i}^{C_{t}} \E({\Xi}) = m C_t$  it follows: 
\begin{multline}
 \mu(t+1) = \E(C_{t+1}) = \E(\E(C_{t+1}|C_{t})) = \\
 \E(m C_t) = m \E(C_{t-1}).
\end{multline}
Utilising this recursiveness, it follows with initial value $C_0 = 1$:
\begin{equation}
  \mu(t) = m^t.
\end{equation}
A similar but more lengthy argument yields (see for example
\citealt*{haccou:05}) or compare to the more complex argument below for \prettyref{eq:var})
\begin{equation}
  \sigma^2(t):= \Var(C_t) = s^2 \frac{m^t (m^t - 1)}{m(m-1)}.
\end{equation}
For $t \gg 0$, this can be approximated as
\begin{equation}
  \sigma^2(t) \approx s^2 m^{2t}
\end{equation}
Hence for a branching process expectation $\mu(t)$ and its standard
deviation $\sigma(t)$ both increase exponentially with $m^t$.
\par
``Individuals'' and ``offspring'' in a branching process as above can
correspond to ``individuals'' in the biological sense but also to
other discrete units that reproduce. In the main text, an individual
is an \ac{AU} that reduplicates with probability $r$ in a time step
and also survives into the next time step with certainty. An
individual \ac{AU} then has either one (itself, no replication) or two
successors (itself and its replication), ie individual reproduction $\Xi$
into the next time step is governed by the following probability
distribution:
\begin{equation}
  \Prob(\Xi = 1) = (1 -r), \quad
  \Prob(\Xi = 2) = r
\end{equation} 
and probabilities for all other outcomes zero. The
individual reproduction process $\Xi$ then has  mean $m = 1+r$ and
variance $s^2 = r (1 -r)$. Note that this branching process cannot die
out ($\Prob(C_t > 0, \forall t) = 1$), as there is at least one successor
for each individual (ie $\Prob(\Xi \geq 1) = 1$). In fact the process
will be non-decreasing $C_{t+1} \geq C_{t}$.

\section{Combined Production Process} \label{apx:full}

The full stochastic production process of \acp{AU} is the combination
of a standard Galton-Watson branching process that governs \ac{AU}
replication and a Bernoulli process (a discrete time Poisson process) that governs \ac{AU} creation. 
\par
Let $C_t$ denote the population size at $t$, and $\Xi$ the individual
\ac{AU} replication process with mean $\E(\Xi) = m$ and variance
$\Var(\Xi) = s^2$ as before. Let further $X_t$ be the Bernoulli creation process with constant
rate $p$, ie $\Prob(X_t = 0) = 1-p$ and $\Prob(X_t = 1) = p$. Then 
$\E(X_t) = p$ and $\Var(X_t) = p(1-p)$ at each time step.  The
process we are interested in is given by the sequence of recursively
defined random variables $C_t$
\begin{equation}
  C_{t+1} = \sum_{i=1}^{C_{t}} \Xi + X_t.
\end{equation}
Utilising independence of the individual $\Xi{}$s and $X_t$ we can
calculate a recursive equation for $\E(C_t)$. First we note that 
\begin{equation}
  \E(C_{t+1}|C_t) 
  =  \sum_{i = 1}^{C_{t}} \E(\Xi) + \E(X_t) 
  =  C_{t} m + p.
\end{equation} Then making use of $\E(A) = \E(\E(A|B)$ it follows:
\begin{equation}
  \E(C_{t+1}) = \E(\E(C_{t+1}|C_t) = m \E(C_t) + p.
\end{equation}
With initial condition $C_0 = \E(C_0) = 0$ this can be resolved in closed
form:
\begin{equation}
  \E(C_{t}) = \frac{p}{m-1} (m^t -1). 
\end{equation}
Consequentially $\E(C_{t})$ increases with $m^t$ for large
$t$. For the variance, we set $Z_t:= \sum_{i=1}^{C_t}$, hence
$C_{t+1} = Z_t + X_t$ 
and  utilising again the independence of $Z_t$ and $X_t$:
\begin{equation}
  \Var(C_{t+1}) = \Var(Z_t) + \Var(X_t)
\end{equation}
and further using the variance partitioning formula $\Var(A) =
\E(\Var(A|B)) + \Var(\E(A|B)$:
\begin{gather}
  \Var(Z_{t}) = \E(\Var(Z_t|C_{t})) + \Var(\E(Z_t|C_t))  \\
  = \E(C_t s^2) + \Var(m C_t) 
  = s^2 \E(C_t) + m^2 \Var(C_t),
\end{gather} and hence we get the recursive formula:
\begin{equation}
  \Var(C_{t+1}) = m^2 \Var(C_t) + s^2 \E(C_t) + \Var(X_t). \label{eq:var} \\ 
\end{equation}
We have already a closed solution for $\E(C_t)$. Inserting
\prettyref{eq:meansol} into \eqref{eq:var} we (or rather \texttt{maxima}
with some subsequent manual simplification) can solve this for
$\Var(C_t)$ with initial condition $\Var(C_0) = 0$ (from $C_0 = 0$):   
\begin{multline}
  \Var(C_t) 
  = m^{2t} \Big( 
    p(1-p) \frac{1-m^{-2t}}{m^2-1} \\
    + p s^2\frac{1 - m^{-t}}{m(m-1)^2}
    - p s^2 \frac{1 - m^{-2t}}{(m-1)(m^2-1)}\Big)
\end{multline}
This is rather cumbersome, and we refrain from further simplification
as it is obvious that for $t \to \infty$ $\Var(C_t) \sim m^{2t}$. As also
this
 $C_t$ is non-decreasing, the same argumentation as in
\prettyref{sec:replication} can be applied to the full process in \prettyref{sec:full}. 
\end{appendix}

\bibliographystyle{elsarticle-harv}
\bibliography{yeast}

\end{document}